\documentclass[a4paper,fleqn]{cas-dc}

\usepackage{lineno,hyperref}
\usepackage[numbers,longnamesfirst]{natbib}
\usepackage{amssymb}

\usepackage{booktabs, makecell, tabularx}

\usepackage{siunitx}%
\usepackage{adjustbox}

\def\tsc#1{\csdef{#1}{\textsc{\lowercase{#1}}\xspace}}
\tsc{WGM}
\tsc{QE}

\begin{document}
\let\WriteBookmarks\relax
\def\floatpagepagefraction{1}
\def\textpagefraction{.001}

\shorttitle{High-precision electron-capture $Q$ value measurement of $^{111}$In for electron-neutrino mass determination}   

\shortauthors{Z.~Ge, T.~Eronen, A.~de~Roubin, K.~S.~Tyrin et al.} 

\title [mode = title]{High-Precision electron-capture $Q$ value measurement of $^{111}$In for electron-neutrino mass determination}




\author[1]{Z.~Ge}[orcid=0000-0001-8586-6134]
\cormark[1]
\fnmark[1]
\ead{z.ge@gsi.de}
\author[1]{T.~Eronen}[orcid=0000-0003-0003-6022]
\ead{tommi.eronen@jyu.fi}
\cormark[1]

\affiliation[1]{organization={Department of Physics, 
                  University of Jyv\"askyl\"a},
                addressline={P.O. Box 35}, 
                postcode={FI-40014}, 
                state={Jyv\"askyl\"a},
                country={Finland}}
\author[2]{A.~de~Roubin}[orcid=0000-0002-6817-7254]
\affiliation[2]{organization={Centre d'Etudes Nucl\'eaires de Bordeaux Gradignan, UMR 5797 CNRS/IN2P3 - Universit\'e de Bordeaux},
                addressline={19 Chemin du Solarium, CS 10120}, 
                postcode={F-33175}, 
                state={Gradignan Cedex},
                country={France}}
                
\author[3]{K.~S.~Tyrin}[orcid=0000-0003-4041-899X]      
\affiliation[3]{organization={National Research Center ``Kurchatov Institute''},
                addressline={Pl. akademika Kurchatova 1}, 
                postcode={123098}, 
                state={Moscow}, 
                country={Russia}}

\author[4]{L.~Canete}[orcid=0000-0002-4229-6386]%
\fnmark[2]
\affiliation[4]{organization={Department of Physics, University of Surrey},
                addressline={Guildford GU2 7XH}, 
                state={Surrey}, 
                 country={United Kingdom}}



\author[1]{S.~Geldhof}[orcid=0000-0002-1335-3505]
\fnmark[3]





\author[1]{A.~Jokinen}[orcid=0000-0002-0451-125X] 
\author[1]{A.~Kankainen}[orcid=0000-0003-1082-7602] 
\author[5]{J.~Kostensalo}[orcid=0000-0001-9883-1256]
\affiliation[5]{organization={Natural Resources Institute Finland},
                addressline={Yliopistokatu 6B}, 
                postcode={FI-80100}, 
                state={Joensuu},
                country={Finland}}
\author[1,6,7]{J.~Kotila}[orcid=0000-0001-9207-5824]
\affiliation[6]{organization={Finnish Institute for Educational Research, University of Jyv\"askyl\"a},
                addressline={P.O. Box 35}, 
                postcode={FI-40014}, 
                state={Jyv\"askyl\"a},
                country={Finland}}
\affiliation[7]{organization={Center for Theoretical Physics, Sloane Physics Laboratory},
                addressline={Yale University}, 
                city={New Haven},
                postcode={Connecticut 06520-8120}, 
                country={USA}} 
\author[3,8]{M.~I.~Krivoruchenko}[orcid=0000-0002-4450-1427]
\cormark[1]
\affiliation[8]{organization={Institute for Theoretical and Experimental Physics, , NRC ``Kurchatov Institute''},
                addressline={B. Cheremushkinskaya 25}, 
                postcode={117218}, 
                state={Moscow}, 
                country={Russia}}
\ead{mikhail.krivoruchenko@itep.ru}
\author[1]{I.~D.~Moore}[orcid=0000-0003-0934-8727] 
\author[1]{D.~A.~Nesterenko}[orcid=0000-0002-6103-2845]    
\author[1]{J.~Suhonen}[orcid=0000-0002-9898-660X]
\author[1]{M.~Vil\'en}[orcid=0000-0002-0375-2502]
\fnmark[4]                
   
\cortext[cor1]{Principal corresponding authors}
\fntext[fn1]{Present address: GSI Helmholtzzentrum f\"ur Schwerionenforschung GmbH, 64291 Darmstadt, Germany}
\fntext[fn2]{Present address: University of Surrey, Department of Physics, Guildford GU2 7XH, United Kingdom}
\fntext[fn3]{Present address: KU Leuven, Instituut voor Kern- en Stralingsfysica, B-3001 Leuven, Belgium}
\fntext[fn4]{Present address: Experimental Physics Department, CERN, CH-1211 Geneva 23, Switzerland}


\begin{abstract}
A precise determination of the ground state $^{111}$In ($9/2^+$) electron capture to ground state of $^{111}$Cd ($1/2^+$) $Q$ value has been performed utilizing the double Penning trap mass spectrometer, JYFLTRAP. A value of 857.63(17) keV was obtained, which is nearly a factor of 20 more precise than the value extracted from the Atomic Mass Evaluation 2020 (AME2020). The high-precision electron-capture $Q$ value  measurement along with the nuclear energy level data of  866.60(6) keV, 864.8(3) keV, 855.6(10) keV, and 853.94(7) keV  for $^{111}$Cd  was used to determine whether the four states are energetically allowed for a potential ultra-low $Q$-value  $\beta^{}$ decay or electron-capture decay. Our results confirm that the excited states of 866.60(6) keV with spin-parity ($J^\pi$) of 3/2$^{+}$ and 864.8(3) keV with $J^\pi$ = 3/2$^{+}$ are ruled out due to their deduced electron-capture $Q$ value  being smaller than 0 keV at the level of around 20$\sigma$  and 50$\sigma$, respectively. 
Electron-capture decays to the excited states at 853.94(7) keV ($J^\pi$ = 7/2$^+$)  and 855.6(10)  keV ($J^\pi$ = 3/2$^+$), are energetically allowed with $Q$ values of 3.69(19) keV and 2.0(10) keV, respectively. The allowed decay transition  $^{111}$In  (9/2$^{+}$) $\rightarrow$ $^{111}$Cd (7/2$^{+}$),  with a $Q$ value of 3.69(19) keV, 
is a potential a new candidate for neutrino-mass measurements by future EC experiments featuring new powerful detection technologies.
{
The results show that the indium level $2p_{1/2}$ for this decay branch leads to a significant increase in the number of EC events in the energy region sensitive to the electron neutrino mass. 
}
 

%

\end{abstract}




\maketitle






\section{Introduction}

The study of rare beta decays is one of the hottest topics in contemporary nuclear physics due the fundamental implications for weak-interaction and neutrino physics. A large amount of information on the relative masses and mixing of neutrinos has been obtained via neutrino-oscillation experiments~\cite{Fukuda1998,SNOCollaboration2002,Gerbino2018a}. The piece which is still missing is the absolute mass scale of the neutrino/antineutrino. Current  experiments to investigate the neutrino/antineutrino mass scale focus on nuclear double-$\beta$  decay~\cite{Suhonen1998,Avignone2008,Ejiri2019} as well as $\beta$ decays of nuclei such as tritium in the KATRIN (KArlsruhe TRitium Neutrino) experiment~\cite{Drexlin2013,Aker2019}, $^{187}$Re  in the MARE (Microcalorimeter Arrays for a Rhenium Experiment) 
experiment~\cite{Nucciotti2012,Ferri2015} and $^{163}$Ho in the ECHo (Electron Capture in $^{163}$Ho)~\cite{Gastaldo2014,Gastaldo2017} and HOLMES~\cite{Alpert2015,Faverzani2016} 
experiments. The former two experiments search for the electron-antineutrino mass while the latter two  search for the electron neutrino mass, both via the slight distortion of the electron end-point spectrum. 
For the beta-decay experiments the fraction of decays in a given energy interval $\Delta{E}$ below the end-point $Q$ value is proportional to $(\Delta{E}/Q)^3$~\cite{Ferri2015}, whereas for the electron capture (EC) this dependence on the $Q$ value can be even more drastic.  Hence, for all these experiments it is essential to have a decay $Q$ value which is as small as possible.
The tritium experiment is based on the $\beta^{}$ transition
$^{3}$H(1/2$^{+}$) $\rightarrow$ $^{3}$He(1/2$^{+}$) which is of the allowed type
(a Fermi and/or Gamow–Teller transition) with a $Q$ value of 18.59201(7) keV~\cite{Myers2015}. The rhenium experiment relies on the $\beta^{}$ transition $^{187}$Re(5/2$^{+}$) $\rightarrow$ $^{187}$Os(1/2$^{-}$) which is of the first-forbidden unique type with the lowest known ground-state-to-ground-state (gs-to-gs) $\beta$-decay $Q$ value of 2.492(30)$_\textrm{stat}$(15)$_\textrm{sys}$  keV~\cite{Basunia2017,Nesterenko2014} (with statistical and systematic uncertainties indicated). 
Owing to the smaller decay $Q$ value and the forbidden nature of the $\beta^-$ transition, the half-life of the rhenium transition is some 9 orders of magnitude longer than that of tritium. On the other hand, the rhenium transition is unique, and thus depends only on one nuclear matrix element (NME) and leads to a universal electron spectral shape like in the case of the allowed decays. The holmium experiment is in a position to use the to-date lowest known gs-to-gs EC transition in the isotope $^{163}$Ho, possessing a $Q$ value of 2.833(30)$_\textrm{stat}$(15)$_\textrm{sys}$ keV~\cite{Eliseev2015}.

The possible existence of other isotopes that could undergo  $\beta$ decay with a low $Q$ value is of significant interest for  future neutrino-mass-scale determination experiments~\cite{Haaranen2013,Suhonen2014,ge2021}. Among cases where $Q$ < 1 keV, these decays have been called “ultra-low”~\cite{Mustonen2010,Mustonen2011,Suhonen2014}. The existence of an ultra-low $Q$ value was first discovered by Cattadori \emph{et al.}~\cite{Cattadori2005}. The intriguing case of the $\beta$ decay of the 9/2$^{+}$ ground state of $^{115}$In to the first excited state  of $^{115}$In (9/2$^{+}$) with a $Q$ value of 0.35(17) keV has been confirmed by a combined effort of HADES underground laboratory and JYFLTRAP Penning trap measurements  with a branching ratio of only 1.07(17)$\times$10$^{-6}$~\cite{Wieslander2009}. The  beta-decay $Q$ value has been further  refined to 0.155(24) keV independently by Mount et al.~\cite{Mount2009}.
The $Q$ value of this branch was recently determined to be 0.147(10) keV with the accurately measured excitation energy of the first excited state of $^{115}$Sn~\cite{Zheltonozhsky_2018}.
A low $Q$-value $\beta^{-}$-decay transition from the  7/2$^{+}$ ground state of $^{135}$Cs to the  second excited state of $^{135}$Ba, with $J^\pi$ = 11/2$^{-}$, was recently investigated at JYFLTRAP. This study was inspired by the theoretical findings reported in \cite{Mustonen2011}.  The measurement confirmed that the decay is energetically allowed  with an ultra-low $Q$ value of 0.44(31) keV,  which  makes this first-forbidden unique transition with a simple universal spectral shape a potential  candidate for antineutrino-mass measurements with almost an  order of magnitude lower $Q$ value than in presently running or planned direct (anti)neutrino-mass experiments~\cite{deRoubin2020}.  
Also, the electron capture of $^{159}$Dy ($3/2^-$) to the $5/2^-$ state in $^{159}$Tb was recently studied at JYFLTRAP~\cite{ge2021159dy}. This decay was revealed to have $Q$ value of 1.18(19)~keV, making it the lowest electron capture $Q$ value known thus far.

In this letter, the gs-to-gs EC decay $Q$ value of  $^{111}$In has been directly  determined with a Penning trap spectrometer for the first time.  The high-precision $Q_{\mathrm{EC}}$ value from this work, combined with the nuclear energy level data  for $^{111}$Cd, was used to examine whether there are excited final states that enable ultra-low $Q$-value EC transitions, and thus could serve as new potential candidates for neutrino-mass measurements.

\section{Experimental method}
The $Q_{\mathrm{EC}}$ value  of $^{111}$In,  the mass difference of the parent nucleus $^{111}$In and its EC  decaying daughter $^{111}$Cd, was measured at the new Ion Guide Isotope Separator On-Line facility (IGISOL) utilizing the JYFLTRAP double Penning trap mass spectrometer~\cite{Eronen2012,Kankainen2020}, at the University of Jyv\"askyl\"a~\cite{Moore2013,Kolhinen2013}.  
A primary beam of  protons with an energy of 40 MeV from the K-130 cyclotron impinged into a naturally abundant indium target with a thickness of a few mg/cm$^2$ at the entrance of the IGISOL gas cell~\cite{Penttila2016}. The secondary products produced from fusion-evaporation reaction were stopped in helium gas and extracted with the help of a sextupole ion guide (SPIG)~\cite{Karvonen2008}, a linear Paul trap with a sextupole electrode configuration. The ion beam with a typical charge state of 1+ is subsequently accelerated to 30 keV and guided to a 55$^\circ$ dipole magnet mass separator with mass resolving power $M/\Delta{M}$ $\approx$ 500, which is sufficient for separation of different isobars in the secondary beam. After the secondary beam has been isobarically separated, the ions of the same mass number (A = 111) including both $^{111}$Cd$^{+}$ and $^{111}$In$^{+}$  are injected into a radiofrequency quadrupole cooler-buncher (RFQ)~\cite{Nieminen2001} and then injected into JYFLTRAP double Penning trap.



JYFLTRAP consists of two cylindrical Penning traps situated inside a 7-T superconducting solenoid. The first trap, performing as the preparation trap, is filled with buffer gas and is used to remove isobaric contaminants via the sideband buffer gas cooling technique~\cite{Savard1991}. This technique alone can usually provide sufficient cleaning with a resolving power of around $10^{5}$ by mass selectively converting ion motion from magnetron to reduced cyclotron motion. 
If a sample requires even higher resolving power for selecting the ions of interest,  the Ramsey cleaning technique~\cite{Eronen2008a} will be employed right after the sideband buffer-gas cooling. 
In this experiment, a purified sample of either $^{111}$In$^{+}$ or $^{111}$Cd$^{+}$ ions was prepared using this high-resolution cleaning technique before their transfer to the second trap to measure the cyclotron frequency:
$\nu_{c}=\frac{1}{2\pi}\frac{qB}{m}$,
where $B$ is the magnetic field strength, $q$ is the charge state and $m$ the mass of the ion.

As both the parent and daughter nuclei are available from the aforementioned fusion-evaporation reaction, the $Q_{\mathrm{EC}}$ value can be determined by measuring the cyclotron frequency ratio $R$ of the parent and daughter ions using an interleaved scanning method. The $Q_{\mathrm{EC}}$ value can be given as the mass difference of the two ion species:
\begin{equation}
\label{eq:Qec}
Q_{\mathrm{EC}}=(M_p - M_d)c^2 = (\frac{\nu_{c,d}}{\nu_{c,m}}-1)(M_d - m_e)c^2+\Delta{\epsilon_{m,d}},
\end{equation}
where $M_p$ and  $M_d$ are the masses of the parent  ($^{111}$In) and daughter  ($^{111}$Cd) atoms, respectively.  ${\nu_{c,d}}/{\nu_{c,m}}$ ($ = R$) is their cyclotron frequency ratio for singly charged ($q = 1$) ions. $m_{e}$ is mass of an electron and $\Delta \epsilon_{m,d}$ the electron binding energy difference of the parent-daughter atoms, which  can be neglected as it is on the order of few eV~\cite{NIST_ASD}. $c$ is the speed of light in vacuum.
Since both the parent and daughter have the same $A/q$, the mass-dependent error effectively becomes inferior compared to typical statistical uncertainty achieved in the measurement. In addition, as the mass difference of the parent and daughter is very small ($\Delta M/M$ < $10^{-5}$), the contribution of uncertainty to the $Q$ value from the mass uncertainty of the reference (daughter), 0.4 keV/c$^2$, can be neglected.

\begin{figure*}[!htb]
\centering
   \includegraphics[width=1.8\columnwidth]{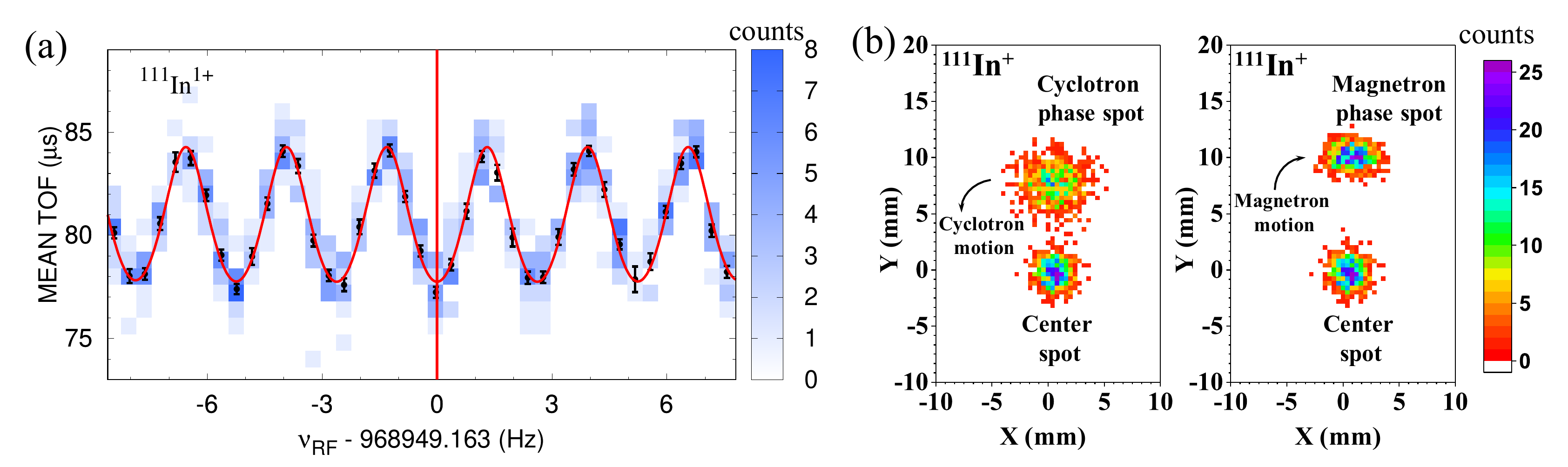}
   \caption{(Color online). (a) A Ramsey time-of-flight ion-cyclotron resonance  (TOF-ICR) for $^{111}$In$^{+}$ acquired with an  25 ms (On) - 350 ms (Off) - 25 ms (On) excitation pattern. The black dots  with uncertainties are the average TOF, and the solid line in red is the fit  to theoretical line shape. The vertical red line demonstrates the central frequency. The color of background shading indicates the
number of ions.
(b) Ion spots (center, cyclotron phase and magnetron phase) of $^{111}$In$^{+}$ on the 2-dimensional position-sensitive MCP detector after a typical PI-ICR excitation pattern with an accumulation time of 400 ms. The magnetron phase spot is displayed on the right side  and the cyclotron phase spot on the left. The angle difference between the two spots relative to the center spot is utilized to deduce the cyclotron frequency of the measured ion species. The number of ions in each pixel is illustrated by color bars.
}  
\label{fig:Ramesy-2-phases}
\end{figure*}

Two methods are utilized to measure the  $\nu_{c}$ of the ions in this work.
The first is  the time-of-flight ion-cyclotron resonance (TOF-ICR) technique~\cite{Koenig1995,Graeff1980}, which has been conventionally used to determine $\nu_{c}$ with either a single quadrupole excitation or the so-called Ramsey excitation~\cite{George2007a,George2007,Kretzschmar2007}. The Ramsey TOF-ICR has a three-fold gain in precision compared to the normal TOF-ICR. For the normal TOF-ICR, once the ions are injected into the precision trap, a short magnetron excitation $\nu_{-}$ is applied followed by a quadrupolar excitation. The frequency of the quadrupolar excitation is scanned around the cyclotron frequency $\nu_c$: $\nu_{c}=\nu_{+} + \nu_{-}$,  where $\nu_{+}$ is the reduced cyclotron frequency and $\nu_{-}$ the magnetron frequency of the ion. Within the excitation time, the motion of the ions in the resonance is fully converted from magnetron to cyclotron with a correctly chosen amplitude. As a result, the ions in resonance undergo a stronger axial force in the magnetic field gradient resulting in a shorter time of flight from the Penning trap to a micro-channel plate (MCP) detector. In this experiment, we used time-separated oscillatory fields for the quadrupolar excitation (Ramsey method). The quadrupolar excitation was applied as two 25-ms fringes separated by 350 ms for $^{111}$In$^{+}$ and  $^{111}$Cd$^+$. An alternated measurement of $\nu_{c}$ of  the parent and daughter ions, $^{111}$In$^{+}$ and $^{111}$Cd$^{+}$, is conducted for every $\sim$ 20 minutes. In total, alternated measurements were performed for $\sim$ 6 hours. A typical Ramsey-type TOF-ICR spectrum obtained for $^{111}$In$^{+}$ is shown in Fig.~\ref{fig:Ramesy-2-phases}(a).

The second method used to measure the $\nu_{c}$ in this experiment was the newly implemented and commissioned  phase-imaging ion-cyclotron-resonance (PI-ICR) technique~\cite{Nesterenko2018}. This technique depends on projecting the ion motion in the Penning trap onto a position-sensitive microchannel-plate (MCP) ion detector and provides around 40 times better resolving power than the TOF-ICR method~\cite{Nesterenko2018,Eliseev2014,Eliseev2013}. Measurement scheme 2 described in~\cite{Eliseev2014} was applied to measure the cyclotron frequency $\nu_{c}$ of the corresponding nuclide. After injection of the ions of interest into the center of the second trap, dipole RF pulses  at the corresponding motion frequencies are used to damp the coherent components of the magnetron and the axial motions. Then  the cyclotron motion is excited at the reduced cyclotron frequency ($\nu_{+}$) to  increase the radius of the corresponding motion to a certain radius in order to set the initial phase of this motion. Afterward, two alternate excitation patterns are applied  to measure the ion cyclotron frequency  $\nu_{c}$.  In the first pattern, the cyclotron motion is converted to the magnetron motion with  a quadrupole excitation  to the same radius and  subsequently the ions perform the magnetron motion for a time t accumulating a certain magnetron phase. After the accumulation time $t$, ions are extracted out of the trap and their position is projected onto a position-sensitive MCP detector~\cite{PS-MCP}. The second pattern is otherwise similar but there the ions first perform  the cyclotron phase for an accumulation time $t$ followed by a consecutive conversion to the magnetron motion.   
The center spot  of the ions of interest is collected by extracting and projecting the ions
onto the MCP detector after injection into the center of the second trap and storing for a few milliseconds without exciting the cyclotron motion. 
The first pattern is utilized to measure the magnetron motion phase whereas the second pattern provides the cyclotron motion phase.
The angle between the positions of two phase images with respect to the center spot is $\alpha_c$ = $\alpha_+ - \alpha_-$, where $\alpha_+$ and $\alpha_-$ are the polar angles of the spot position on the detector for the magnetron and cyclotron motion phases. The cyclotron frequency $\nu_{c}$  is deduced from: 
\begin{equation}
\label{eq:nuc2}
\nu_{c}=\frac{\alpha_{c}+2\pi n_{c}}{2\pi{t}},
\end{equation}
where $n_{c}$ is the number of integer cyclotron revolutions of the measured ion during the phase accumulation time $t$. The accumulation time $t$ in this measurement for $^{111}$In$^{+}$ and  $^{111}$Cd$^{+}$ was 400 ms. 
To obtain the $\nu_{c}$, 
the patterns were repeatedly measured, each lasting approximately 1 minute. In between, the center spot was also recorded. 
The positions of the magnetron-motion and cyclotron-motion phase spots were chosen such that the angle $\alpha_c$  did not exceed a few degrees. This reduces the shift in the frequency ratio $R$ between the $^{111}$In$^{+}$ and  $^{111}$Cd$^{+}$ ions due to the conversion of the cyclotron motion to magnetron motion and any possible distortion of the ion-motion projection onto the detector to a level well below 10$^{-10}$~\cite{Eliseev2014}. A measurement of the cyclotron  and magnetron  phase spots 
is shown in the left and right panels of Fig.~\ref{fig:Ramesy-2-phases}(b), respectively. The starting delay of the excitation of the cyclotron motion was scanned over one magnetron period and the extraction delay after the conversion excitation was varied over one cyclotron period to account for any residual magnetron and cyclotron motion that could shift the different spots. 
In total the data were accumulated for around 26 hours.



\begin{table}[!htb]
 \fontsize{8}{9}\selectfont
\caption{Potential candidates for ultra-low $Q$-value transitions from the $^{111}$In  (9/2$^{+}$) ground state to the excited states in $^{111}$Cd. The first column gives the excited final state in $^{111}$Cd and the second column gives the decay type to this state.  The third column gives the derived decay $Q_{\mathrm{EC}}$ value, in units of keV,  from the literature (AME2020)~\cite{Wang2021} and the fourth column gives the corresponding value as derived from this work. The  last  column gives the experimental excitation energy $E^{*}$ with the experimental error~\cite{NNDC} in units of keV. The symbol 2nd FU stands for second-forbidden unique.
   }
     \begin{tabular*}{0.48\textwidth}{@{}ccccc@{}}
  \toprule
 Final state & Decay type & \makecell[c]{$Q_{\mathrm{EC}}$ \\AME2020} &\makecell[c]{$Q_{\mathrm{EC}}$ \\ This work} & $E^{*}$ \\
   \midrule
   $^{111}$Cd (7/2$^{+}$)&   allowed  &   6.3(34) &3.69(19) & 853.94(7) \\
   $^{111}$Cd (3/2$^{+}$)& 2nd FU & 4.6(36) &2.0(10)   &855.6(10) \\
   $^{111}$Cd (3/2$^{+}$)&2nd FU &   -4.6(35)&-7.17(35)  &864.8(3)  \\
   $^{111}$Cd (3/2$^{+}$)&  2nd FU &   -6.4(34)&-8.97(18)    &866.60(6) \\
     \bottomrule
      \end{tabular*}
   \label{table:low-Q}
\end{table}


\section{$Q$-value determination}
\begin{figure}[!htb]
   \includegraphics[width=0.95\columnwidth]{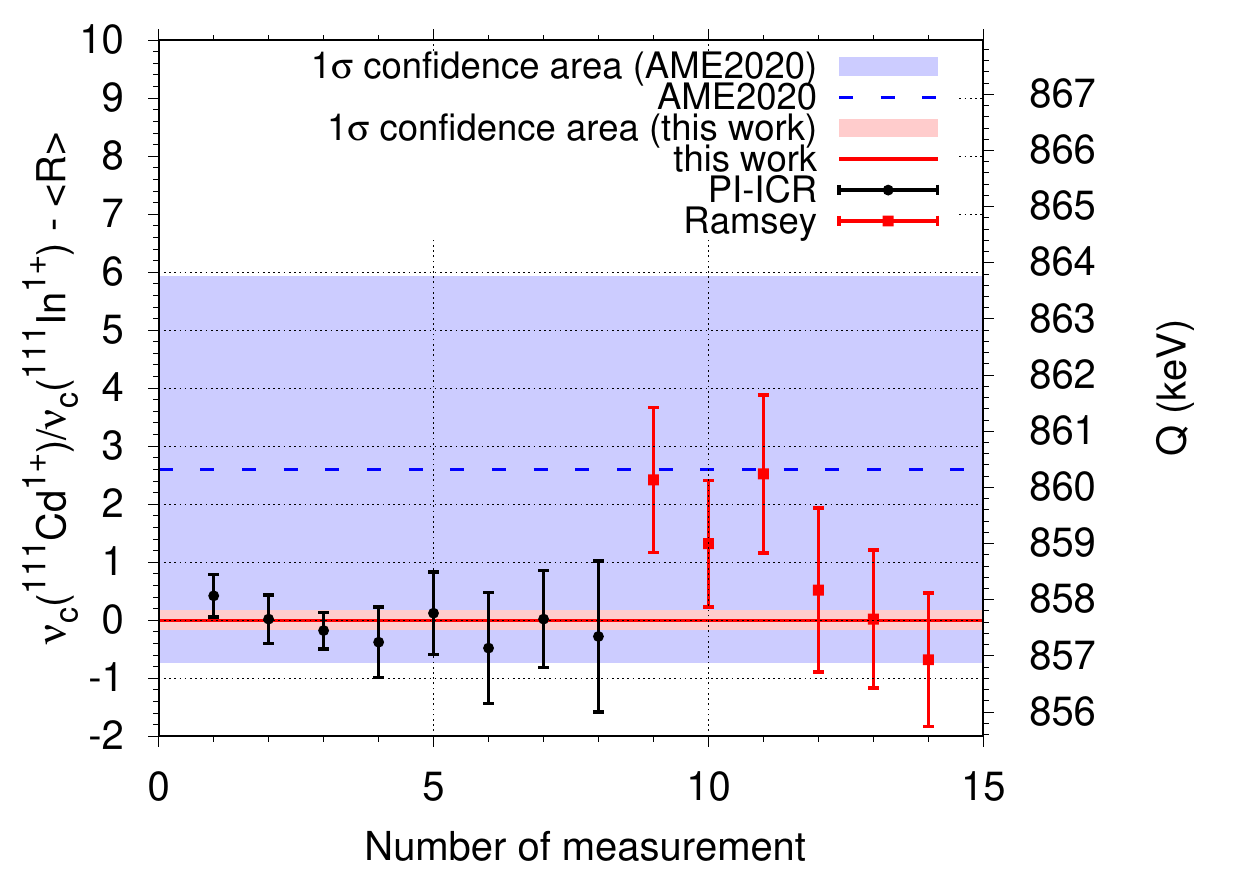}
   \caption{(Color online). The measured (left axis) cyclotron frequency ratios  $R$ ( $\nu_c$($^{111}$Cd$^{+}$)/$\nu_c$($^{111}$In$^{+}$)) and (right axis)) $Q$-values in this work compared to values adopted from AME2020~\cite{Wang2021}. Data shown as black dots and red squares with uncertainties are measured with PI-ICR and TOF-ICR, respectively. The weighted average value from this work  $\langle R \rangle$ = 1.000 008 301 9(17) is represented by the solid red line and its 1$\sigma$ uncertainty band is shaded in red. The dashed blue line represents the 
   AME2020 value with its 1$\sigma$  uncertainty area shaded in blue.}
   \label{fig:ratio}
\end{figure}

The TOF-ICR and PI-ICR data were split to 6 and 8 parts for final fitting, respectively.  The cyclotron frequency $\nu_{c}$ of the daughter $^{111}$In$^{+}$ as a reference was  linearly interpolated to the time of the measurement of the parent $^{111}$Cd$^{+}$ (ion of interest) to deduce the cyclotron frequency ratio $R$. Bunches with up to five detected ions per cycle were  considered in the data analysis in order to reduce a possible cyclotron frequency shift due to ion-ion interactions~\cite{Kellerbauer2003,Roux2013}. The count-rate related frequency shifts were not observed in the analysis of both methods. Contribution of  temporal fluctuations of the magnetic field  to the final frequency ratio uncertainty was less than 10$^{-10}$ since the parent-daughter measurements were interleaved < 25 minutes for the TOF-ICR measurement and < 5 minutes for the PI-ICR measurement. The frequency shifts in the PI-ICR measurement due to ion image distortions, were well below the statistical uncertainty, and thus were ignored in the calculation of the final uncertainty.  The weighted mean ratio $\langle R \rangle$ of the single ratios for both  TOF-ICR and PI-ICR data was calculated along with the inner and outer errors.  The maximum of the inner and outer errors of the ratios were taken as the weights to calculate $\langle R \rangle$.  The Birge ratio~\cite{Birge1932} of the outer and inner errors is 0.915 and the inner error is taken as the final error of this work. In Fig.~\ref{fig:ratio}, results of the analysis including all data from both Ramsey-type TOF-ICR and PI-ICR measurements with comparison to literature values are shown. The final frequency ratio with its  uncertainty as well as the corresponding $Q$ value are $ \langle R \rangle $=  1.000 008 301 9(17) and $Q_{\mathrm{EC}}$ = 857.63(17) eV, respectively.

\begin{figure}[!htb]
   \includegraphics[width=1.0\columnwidth]{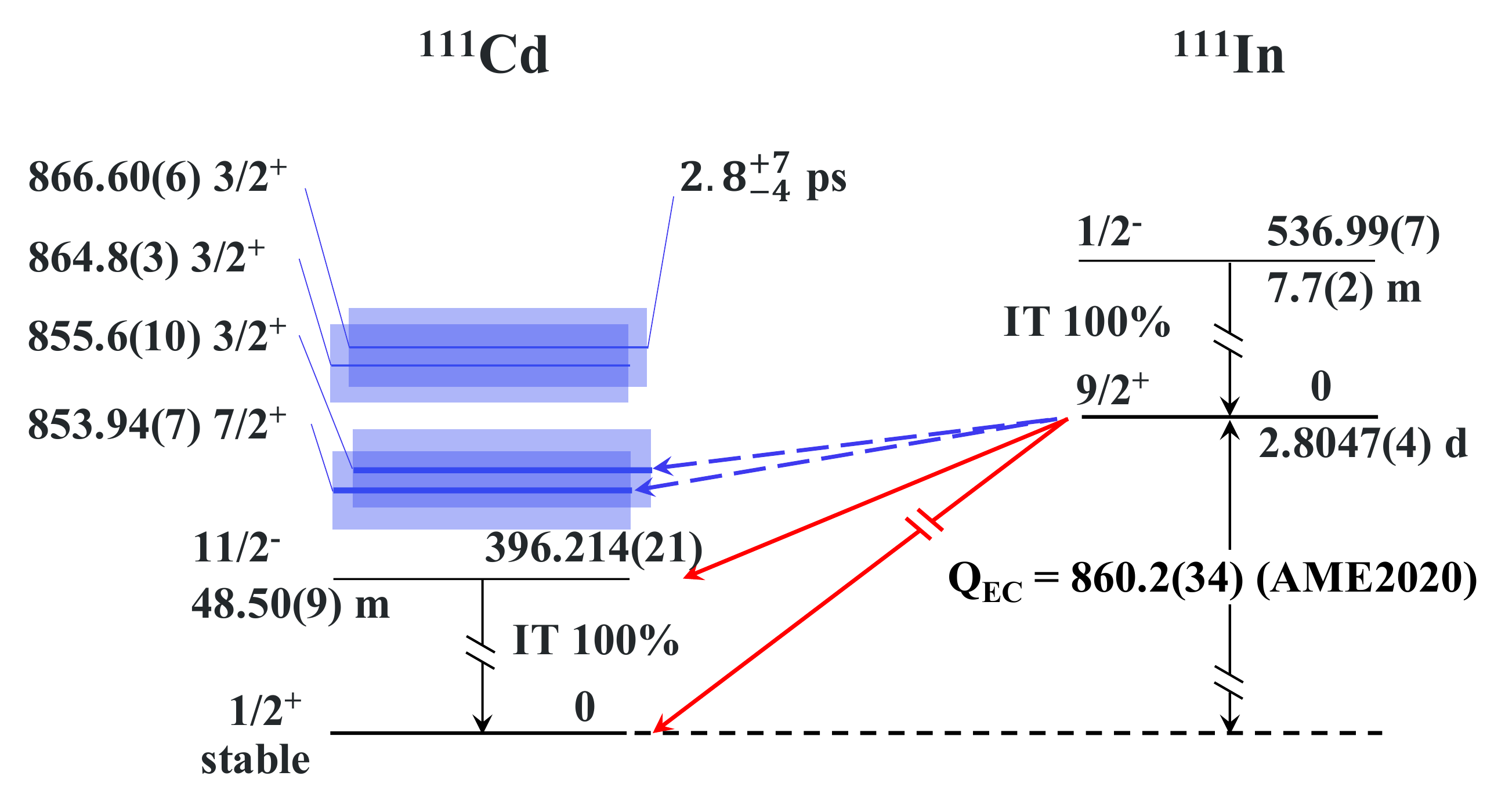}
   \caption{(Color online). EC-decay scheme for the gs-to-gs/excited-state transitions from $^{111}$In to states of $^{111}$Cd.  Possible ultra-low $Q$-value decay branches are indicated by dashed blue lines. The shaded areas in blue illustrate the corresponding 1$\sigma$ uncertainties of the $Q$ values. The data for this level scheme are adopted from~\cite{Huang2021,Wang2021,NNDC}. The decay $Q$ values and excitation energies are in units of keV.}
   \label{fig:level-scheme}
\end{figure}

The mass-excess of the parent nucleus $^{111}$In (9/2$^{+}$) with a half-life of 2.8047(4) days~\cite{NNDC} was deduced to be  -88394.57(42) keV/c$^2$, for which the uncertainty mainly comes from the uncertainty (0.4 keV/c$^2$) of its ground-state daughter $^{111}$Cd (1/2$^{+}$) as a mass reference. The  $Q_{\mathrm{EC}}$ = 857.63(17) keV from this work  is a factor of nearly 20 times more precise than that derived from the evaluated masses  in AME2020~\cite{Huang2021,Wang2021}. The measured  $Q_{\mathrm{EC}}$ value has a deviation of -2.57 keV from the AME2020 value,  860.2(34) keV.  The  value in AME2020 was derived primarily from the difference between the atomic mass of the parent $^{111}$In and that of the daughter $^{111}$Cd species as listed therein. The AME2020 mass value for $^{111}$Cd has been derived from two nuclear reactions  $^{111}$Cd(n,$\gamma$)$^{112}$Cd and $^{110}$Cd(n,$\gamma$)$^{111}$Cd with mass-determination influences on the primary nuclide of 80.7~$\%$ and 19.3~$\%$, respectively.  The AME2020 mass value for  $^{111}$In was principally based on three reaction measurements $^{113}$In(p,t)$^{111}$In-$^{112}$Cd(p,t)$^{110}$Cd with an influence of 69.0 $\%$,  $^{108}$Cd($^3$He,d)$^{109}$In-$^{110}$Cd($^3$He,d)$^{110}$In with  an influence of 19.3 $\%$, $^{113}$In(p,t)$^{111}$In-$^{115}$In(p,t)$^{113}$In with  an influence of 11.7  $\%$. Thus, the contributions  to the mass uncertainty of both the parent $^{111}$In and daughter $^{111}$Cd nuclei come from indirect measurements in AME2020. Previous studies have already demonstrated that mass values derived via indirect methods, such as decay spectroscopy and nuclear reactions, can have large discrepancies with those from direct mass measurements and be inaccurate over a broad range of mass numbers~\cite{Hardy1977,Eliseev2011a,Nesterenko2019}. The high-precision EC energy from this work, together with the nuclear energy level data (see Table~\ref{table:low-Q}) 
from~\cite{NNDC} of the excited states of $^{111}$Cd, was used to determine the $Q_{\mathrm{EC}}$ value  of these four states, see Fig~\ref {fig:level-scheme}. The calculated $Q_{\mathrm{EC}}$ values of the potential candidate transitions from  the  ground state of the parent $^{111}$In  to the excited states of the daughter  $^{111}$Cd are tabulated in Table~\ref{table:low-Q}. Our results confirm that the decay transitions from the ground state of $^{111}$In to the two excited states with excitation energies of 866.60(6) keV and 864.8(3) keV are energetically forbidden. The corresponding $Q_{\mathrm{EC}}$ values of  -7.17(35) keV and -8.97(18) keV are negative at levels of around 20$\sigma$  and 50$\sigma$. 
{Contrary to this, the two excited states 
at 855.6(10)  keV with $J^\pi$ = 3/2$^{+}$ and 
at 853.94(7) keV with  $J^\pi$ = 7/2$^{+}$
correspond  to  the positive $Q_{\mathrm{EC}}$ values of 2.0(10) keV and 3.69(19) keV, respectively.}

The energy of the emitted neutrino, $E_{\nu}$, is concentrated around  
$E_{\nu x} = Q_{\mathrm{EC}} - \epsilon_{fx}$,
where $\epsilon_{fx}$ is the binding energy of the captured electron corresponding to the $x$-th atomic shell of the daughter nucleus.  The tiny recoil energy of the nucleus is neglected in this expression. 
Only $s$ and $p_{1/2}$ electrons from the third and higher shells can be captured due to the angular-momentum conservation and owing to the finite overlap of their wave functions with the nucleus.
{
The binding energies of the K and L1 electron shells~\cite{Larkins1977} exceed the $Q_{\mathrm{EC}}$ values of 2.0(10) keV and 3.69(19) keV, and thus the captures from the K and L1 shells are excluded. In the 2nd FU transition, EC to the L2 level is excluded, whereas in the allowed transition the L2 level is at the boundary of the physical region.
This leaves captures from the atomic shells M1, M2, N1, N2, O1, O2, and possibly L2 allowed.
}
We deduce the average neutrino energies $E_{\nu x}$ of 2.92(19) keV for the 7/2$^{+}$ state of the daughter and 1.3(10) keV for the 3/2$^{+}$ state of the daughter in the case of EC from the M1-atomic shell with $\epsilon_{fx}$ = 770.2 eV~\cite{Larkins1977}. The average energy $E_{\nu x}$ of the emitted neutrino should be as small as possible in order to be a preferable candidate for the neutrino-mass determination.  Here the smallest possible value is 1.3(10) keV, corresponding to the transition to the 855.6(10)  keV 3/2$^+$ state, but it proceeds via a 2nd FU nuclear transition and is thus excluded from considerations for future neutrino-mass measurements due to the associated long half-life (see below). 
In any case, in order to determine whether this transition is associated with an ultra-low $Q$ value  (< 1 keV), more precise energy level data is required.
The allowed decay transition  $^{111}$In   (9/2$^{+}$) $\rightarrow$ $^{111}$Cd (7/2$^{+}$),  with a $Q_{\mathrm{EC}}$ value of 3.69(19) keV, on the other hand, has the neutrino energy $E_{\nu x}=2.92(19)$ keV for the electron  M1-capture and a promisingly short decay half-life (see below) in order to become a potential candidate for future neutrino-mass measurements.

\section{EC rate estimates and 
energy spectrum}

The main observable of EC is the decay constant $\lambda_{x}$, which determines the probability to capture an electron with quantum numbers $x=(njl)$ per unit time.
The following formula for $\lambda_{x}$ can be derived from the standard $\beta$-decay Hamiltonian in the zero-width limit of the excited electron shells \cite{Behrens1982}:
\begin{equation}
\label{eq:lambda}
\lambda_x = \frac{G_{\beta}^2}{(2\pi)^2}n_x \mathcal{B}_x\beta^2_x E_{\nu x} p_{\nu}(E_{\nu x}) C_x,
\end{equation}
where $G_{\beta}=G_{F}V_{ud}$, $G_{F}$ is the Fermi constant, $V_{ud}$ is the mixing coefficient of the $u$- and $d$-quarks, $n_{x}$ is the relative occupation number of the $x$ shell, $p_{\nu}(E_{\nu}) = \sqrt{E_{\nu}^2 + m_{\nu}^2}$ is the momentum and $m_{\nu}$ is the effective mass of the electron neutrino.
The quantity $\beta_x$ is a numerical constant, proportional to the value of the wave function of the captured electron of the parent atom inside the parent nucleus. The overlap-exchange correction factor $\mathcal{B}_x$ is a square of the atomic matrix element which takes into account many-body effects in the atomic shells. The phase-space volume proportional to $E_{\nu x}p_{\nu}(E_{\nu x})$ neglects the nuclear recoil. The factor $C_{x}$ is a shape function of the transition; it contains information about the nuclear structure including the nuclear matrix element and 
the coefficients associated with transformation properties of the decay amplitude. 
To calculate values of $\mathcal{B}_x$ and $\beta_x$, electron wave functions of multi-electron atoms are required. These wave functions are found by the Dirac-Hartree-Fock method with the use of a package for calculations of the relativistic atomic shell structure G\textsc{RASP}2018 \cite{GRASP2018}. 
We compute radial wave functions for the ground-state electron configuration of indium.
To determine the values of $\beta_x$, the radial wave function of the electron obtained numerically is compared with the analytical decomposition of the bound-state wave function of the Dirac equation inside the nucleus: 
\begin{equation}
\label{eq:fx}
f^{(\pm)}_{x}(r)=\beta_{x}\frac{(p_{x}r)^{k-1}}{(2k-1)!!}\sum_{s=0}^{\infty}c^{(\pm)}_{xs}\left(\frac{r}{R_{\mathrm{nucl.}}}\right)^s,
\end{equation}
where $k=|\kappa|$, $\kappa = - (j - l)(2j + 1)$, and $\hbar = c = m_e = 1$ is assumed. Functions $f^{(\pm)}_x(r)$ are upper and lower radial components of the Dirac bispinor. The  
parameter $p_{x}=\sqrt{1-(1-\epsilon_{ix})^2}$ is determined by the binding energy $\epsilon_{ix}$ of an electron in the parent atom. 
The combination $\beta_x p_x^{k-1}$ is usually extracted from numerical calculations, because  $p_x^{k-1}$ enters the decay probability as a multiplier of $\beta_x$.
We use the root-mean-square radius of $R_{\mathrm{nucl.}}=4.587$ fm for the $^{111}$In nucleus. The expansion coefficients $c^{(\pm)}_{xs}$ decrease rapidly for finite-size nuclear potentials, particularly for the Fermi distribution of nuclear charge used for modeling the nuclear potential in G\textsc{RASP}2018. 
We extract $\beta_x p_x^{k-1}$ using the first term of the series (\ref{eq:fx}) with
$c_{x0}^{(\pm)} = ( k \mp \kappa)/(2k)$; the radial wave functions $f^{(\pm)}_{x}(r)$ are evaluated at the first non-zero interpolation point $r = 1.1 \times 10^{-4}$ fm.   

A complete description of the EC process requires the inclusion of many-body effects associated with electrons from the atomic shells of the initial and final atoms. 
Since the nuclear charges and the numbers of electrons in the initial and final states are different, the corresponding wave functions of electrons with different $x$ are not orthogonal, and the overlap of wave functions with the same $x$ is not perfect.
This leads to shaking effects, i.e., excitations and autoionization of the final state of the atom.
Besides shaking effects, many-body correlations lead to a modification of the transition rate, which can be expressed as an exchange-and-overlap correction factor $\mathcal{B}_x$ \cite{Bambynek1977,Behrens1982}. For calculation of the exchange-and-overlap correction factor $\mathcal{B}_x$ we follow the Vatai approach \cite{Vatai:1970ldf} where only diagonal and linear off-diagonal elements of the transition amplitude are taken into account. 
In this approach all the electrons, with the exception of the captured electron, retain their quantum numbers from the parent atom. The corresponding sets of electron wave functions of the final atom are calculated numerically. Shaking effects of the final atom are not taken into account. The values $\beta_x p_x^{k-1}$ and $\mathcal{B}_x$ are shown in Table \ref{table:bB}. The results are in good agreement with the values reported in Ref. \cite{Bambynek1977}. The exchange-and-overlap correction factor enhances the EC rates from the mid and upper shells.

\begin{table*}[!htb]
 \fontsize{8}{9}\selectfont
   \caption
   {
   Parameters $\beta_x p_x^{k-1}$ defined in Eq.~(\ref{eq:fx}) and exchange-and-overlap correction factors $\mathcal{B}_x$ for indium to cadmium EC are shown in units $\hbar=c=m_{e}=1$. 
   The last lines show EC partial half-lives $t^{(x)}_{1/2}$ for capture from the shells $x$ in units of years for the allowed $^{111}$In$(9/2^{+}) \rightarrow ^{111}$Cd$(7/2^{+})$ and second-forbidden unique $^{111}$In$(9/2^{+}) \rightarrow ^{111}$Cd$(3/2^{+})$ transitions. Captures from the M1 and N1 shells are dominant for the allowed transition. In the case of the 2nd FU transition, the capture from the M5 shell is the most probable. The electron shell structure is modeled using the software package GRASP2018 \cite{GRASP2018}.
   }
    \begin{tabular*}{1.00\textwidth}{@{}ccccccccc@{}}
   \toprule
$x$ &K &L1 &L2 &L3 &M1 &M2 &M3 &M4
\\ 
   \midrule
$\beta_x p_x^{k-1}$ &6.094$\times$10$^{-1}$ &2.081$\times$10$^{-1}$ &3.136$\times$10$^{-2}$ &4.948$\times$10$^{-2}$ &9.147$\times$10$^{-2}$&1.433$\times$10$^{-2}$&2.279$\times$10$^{-2}$&1.052$\times$10$^{-3}$ \\
$\mathcal{B}_x$ &0.987 &1.034 &0.967 &0.966 &1.089 &1.015 &1.010 &0.960  \\
Allowed, $t^{(x)}_{1/2}$ [yr] &&&& &5.14$\times$10$^{4}$ &2.07$\times$10$^{6}$ &   &   \\
2nd FU, $t^{(x)}_{1/2}$ [yr] &&&&
& $1.30 \times 10^{25}$ 
& $3.25 \times 10^{26}$ 
& $2.44 \times 10^{20}$ 
& $6.92 \times 10^{22}$ 
\\
   \toprule
$x$ &M5 &N1 &N2 &N3 &N4 &N5 &O1 &O2
\\ 
   \midrule
$\beta_x p_x^{k-1}$ &1.922$\times$10$^{-3}$  &3.929$\times$10$^{-2}$ &5.874$\times$10$^{-3}$ &9.323$\times$10$^{-3}$ &3.748$\times$10$^{-4}$ &6.807$\times$10$^{-4}$&1.093$\times$10$^{-2}$&1.186$\times$10$^{-3}$\\
$\mathcal{B}_x$ &0.939 &1.160 &1.092 &1.098 &1.014 &1.021 &1.256 &1.247  \\  
Allowed, $t^{(x)}_{1/2}$ [yr] &
& $1.74 \times 10^{5}$  & $8.08 \times 10 ^{6}$ &  &  & & $1.95 \times 10 ^{6}$ & $3.34 \times 10 ^{8}$\\
2nd FU, $t^{(x)}_{1/2}$ [yr] 
& $6.78 \times 10^{17}$ 
& $4.97 \times 10^{24}$ 
& $2.08 \times 10^{26}$ 
& $3.52 \times 10^{20}$ 
& $2.10 \times 10^{23}$ 
& $3.19 \times 10^{18}$ 
& $4.28 \times 10^{25}$ 
& $7.32 \times 10^{27}$ \\
  \bottomrule
   \end{tabular*}
   \label{table:bB}
\end{table*}


The allowed nuclear transition is accompanied by a change in the spin of the nucleus by one unit without a change in parity. This is a Gamow-Teller (GT) transition, and its shape function $C_x$ is a constant proportional to the GT nuclear matrix element: 
\begin{equation}
\label{eq:Cx}
C_x= ~(^{\mathrm{A}}F^{(0)}_{101})^2,
\end{equation}
where
$^{\mathrm{A}}F^{(0)}_{101} = {g_{\rm A}} M_{\rm GT} /{\sqrt{2J_i+1}}$. 
The value of the GT matrix element was computed in the nuclear shell-model (NSM) framework using the shell model code NuShellX @MSU~\cite{nushellx}. The effective Hamiltonian jj45pnb~\cite{jj45pnb} was used in a model space consisting of the proton orbitals $0f_{5/2}-1p-0g_{9/2}$ and neutron orbitals $0g_{7/2}-1d-2s$. The neutron orbital $0h_{11/2}$ was left empty in order to relax the formidable computational burden. With an unquenched value $g_{A}=1.27$ of the axial coupling this leads to $^{A}F^{(0)}_{101}=0.09$. The structure of $^{111}$In and $^{111}$Cd isotopes was also studied
by means of the microscopic interacting boson-fermion model
(IBFM-2) \cite{iachello_isacker_1991}, which is an extension of the interacting boson model (IBM-2) \cite{iachello_arima_1987} to study odd-mass nuclei.
In the IBFM-2 $\beta$ decays are described in terms of one-nucleon
transfer operators, which are obtained following the method
that avoids the use of the number-operator approximation, as
discussed in \cite{PhysRevC.93.034332, PhysRevC.95.034317}.
IBM-2 parameters for the even-even core $^{110}$Cd and $^{112}$Sn nuclei were taken from Ref. \cite{PhysRevC.44.1508} or were fitted to reproduce the spectroscopic data of the low lying energy states, respectively. The unperturbed single-particle energies for proton $0f_{5/2}-1p-0g_{9/2}$ and neutron $0g_{7/2}-1d-2s-0h_{11/2}$ orbitals were taken from \cite{PhysRevC.94.034320}.  Finally, the adopted boson-fermion interaction parameters for negative-parity states read as $\Gamma$ = -0.05 and
 $A$ = -0.22 for $^{111}$In, and $\Gamma$ = 0.75, $\Lambda$ = -1.00 and $A$ = 0.30 for $^{111}$Cd. With an unquenched value $g_{A}=1.27$ of the axial coupling this leads to $^{A}F^{(0)}_{101}=0.08$, quite close to the NSM result. 

In allowed transitions, the $s$ and $p_{1/2}$ shells of electrons with $k=1$ give dominant contributions to EC. The Q value imposes an additional constraint, and only electrons with $\epsilon_{fx} < Q_{\mathrm{EC}}$ can be captured, thus forcing the reaction to start from 
the $3s$ shell. Due to the uncertainty of $190$ eV in the Q value, the $2p_{1/2}$ level with a binding energy of $\epsilon_{fx} = 3727$ eV can also contribute to the transition. This level is located above the endpoint by $37 \pm 190$ eV. 

{
Unfortunately, the accuracy of measurement of the Q value does not allow to make an unambiguous conclusion about the position of the $2p_{1/2}$ level relative to the endpoint. Here we consider two options: the subthreshold level for $Q_{\mathrm{EC}} = 2690$ eV and the resonance level for $Q_{\mathrm{EC}} = 2730$ eV. These Q values agree with each other within the experimental error, however, in relation to the energy spectrum near the endpoint the physical consequences turn out to be quite different.
}


{Calculations for $Q_{\mathrm{EC}} = 2690$ eV lead to a total half-life of 
$t_{1/2} = 3.8 \times 10^{4}$ years. Table \ref{table:bB} shows half-lives of the specific dominant channels. The experimental error in $Q_{\mathrm{EC}}$ introduces about 10\% uncertainty in these estimates. The possible contribution of the $2p_{1/2}$ level to the total probability is small due to the smallness of the phase-space volume near the endpoint. For $Q_{\mathrm{EC}} = 3730$ eV and $m_{\nu} = 0$ the partial half-life of the level can be estimated to be $t_{1/2} = 4.8 \times 10^{11}$ years.
}

\begin{figure}[t]
\begin{center}
\includegraphics[angle = 0,width=0.43\textwidth]{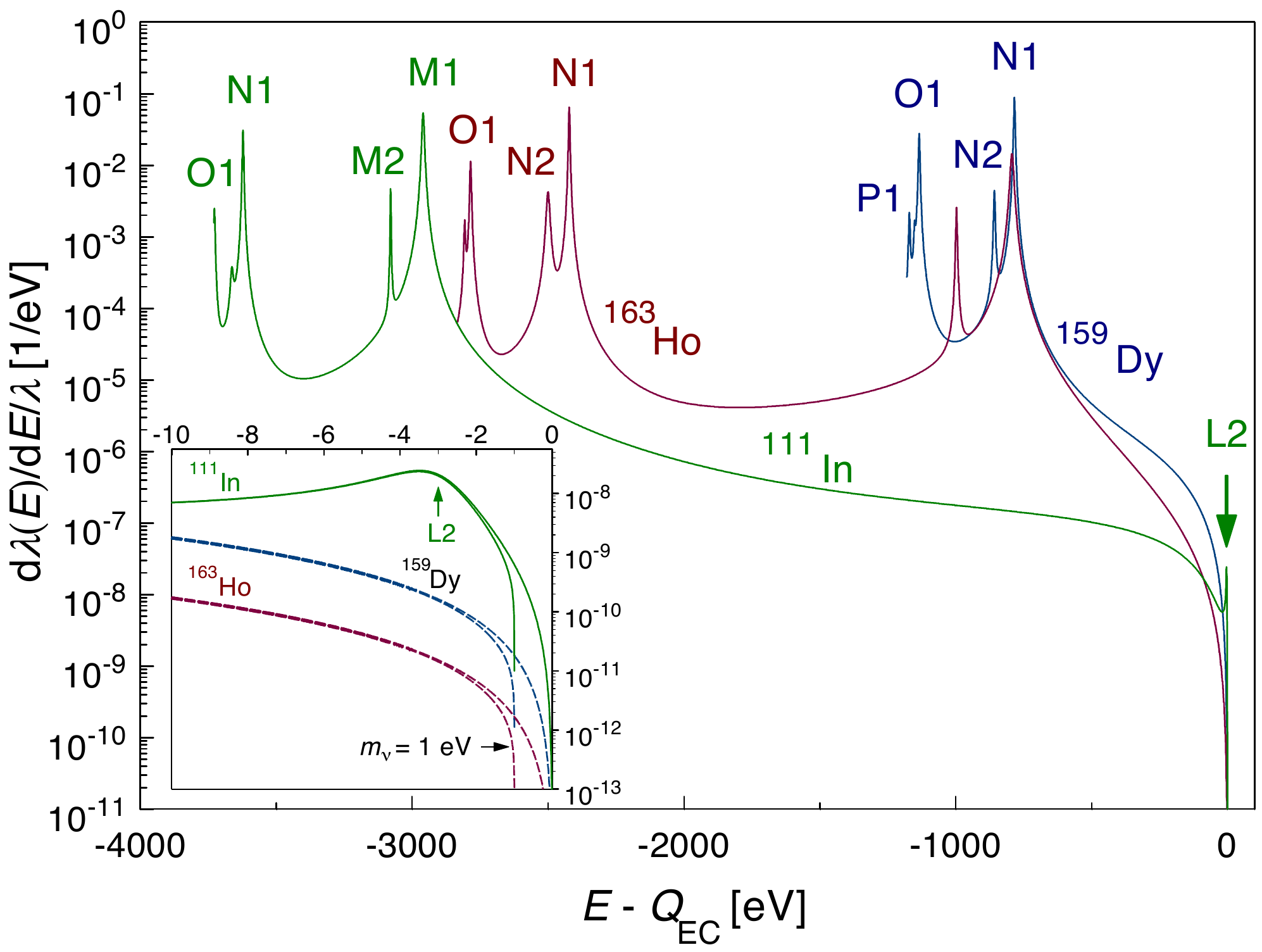}
\end{center}
\vspace{-6mm}
\caption{(Color online)
{
Normalized energy spectrum for $Q_{\mathrm{EC}} = 3730$ eV as a function of the de-excitation energy, $E$, in the EC process $^{111}$In(9/2$^{+}) \rightarrow ^{111}$Cd$ (7/2^{+})$. Symbols M1, M2, N1, O1 indicate electron holes formed in the $^{111}$Cd atom; the holes N2 and O2, which are faintly distinguishable, are not indicated. The EC energy spectra in $^{159}$Dy and $^{163}$Ho atoms are from Refs. ~\cite{ge2021b,Gastaldo2017}. For the considered value of $Q_{\mathrm{EC}}$, the level $2p_{1/2}$ shows up in the physical region as a narrow peak with a width of 2.42 eV \cite{CAMPBELL20011} 
at a distance of 3 eV from the endpoint. This level, L2, significantly increases the EC counting rate in the neutrino-mass sensitive region, as shown on an enlarged scale in the inset in the left lower corner of the figure. }
}
\label{ECplot40}
\end{figure}

The forbidden transition is accompanied by a change in the nuclear spin by three units without a change in parity. This is the 2nd FU transition; its shape function has the  form:
\begin{equation}
\label{eq:Cx2}
C_x=\frac{8 p^{2k-2}_{x} p_{\nu}^{6-2k}(E_{\nu x}) R_{\mathrm{nucl.}}^4 }{15 (2 k-1)! (7 -  2k)! }~(^{A}F^{(0)}_{321})^2.
\end{equation}
The total half-life of the $^{111}$In (9/2$^{+}$) $\rightarrow$ $^{111}$Cd (3/2$^{+}$) decay with the shell-model estimate $^{A}F^{(0)}_{321} = - 0.25$ can be found to be $t_{1/2} = 5.6 \times 10^{17}$ years, thereby excluding this transition from the set of possible candidates for electron-neutrino mass measurements. The normalized partial half-lives for the particular channels are given in Table \ref{table:bB}. The calculations use the electron binding energies tabulated by Larkins \cite{Larkins1977}. 


The energy spectrum of the allowed decay is constructed with the use of Eq.~(\ref{eq:lambda}) 
taking into account the finite decay width of the excited atomic shells:
\begin{equation}
\frac{\mathrm{d}\lambda(E)} {\mathrm{d}E}= \frac{G_{\beta}^{2}}{(2\pi)^{3}}\sum_{x}
 \frac{n_{x} \mathcal{B}_{x}\beta_{x}^{2} E_{\nu} p_{\nu}(E_{\nu}) C_{x} \Gamma_{fx}
}{(E_{\nu} - E_{\nu x})^{2}+\Gamma_{fx}^{2}/4},
\label{spec}
\end{equation}
where $E = Q_{\mathrm{EC}}-E_{\nu}$ is the de-excitation energy of the final atomic shells and $\Gamma_{fx}$ is the electromagnetic decay width of the vacancy $x$ in the daughter atom. 
An incoherent sum of the contributions of individual orbitals is taken, the level widths are adopted from \cite{CAMPBELL20011}. The possible dependence of $\Gamma_{fx}$ on energy is neglected. The widths of the O1 and O2 levels are not available, and we assume $\Gamma_{f\text{O1}} = \Gamma_{f\text{O2}} = 5$ eV. The EC is accompanied by multiple excitation of higher orbitals, as well as knocking out of electrons into the continuum, which creates a fine structure of the energy spectrum \cite{Robertson:2014fka,Faessler:2015txa,Faessler:2015pka,Faessler:2014xpa,Faessler:2016hxd,DeRujula:2016fdu}. These structures are seen experimentally in the EC on the holmium atom and are well described theoretically~\cite{Brass:2017kov,Brab:2020uzx}.
The total decay rate in the zero-width approximation, $\lambda = \sum \lambda_x$, differs from the integral of the expression (\ref{spec}) by no more than a percent. 

{
Figure \ref{ECplot40} shows the normalized EC energy spectrum as a function of the energy, $E$, deposited in a calorimeter through de-excitation of the atomic shells for $Q_{\mathrm{EC}} = 3730$ eV. The energy $E_{\nu}$ is carried away by the neutrino without interactions. For comparison, the EC spectra in $^{159}$Dy and $^{163}$Ho are shown~\cite{ge2021b,Gastaldo2017}. The spectra are normalized to a unit area. The spectra are sensitive to the electron neutrino mass at $E \sim Q_{\mathrm{EC}}$. This region is shown on an enlarged scale in the lower left corner of the figure; the spectra are given for the neutrino masses of 0 and 1 eV.
The region adjacent to the endpoint is highly sensitive to small variations of the Q value because the $2p_{1/2}$ level, L2, can enter the physical region and can manifest itself as a resonance. The $2p_{1/2}$ level radically increases the number of recorded events near the endpoint. In this respect, for $Q_{\mathrm{EC}} = 3730$ eV, indium is about two orders of magnitude more effective than dysprosium and three orders of magnitude more effective than holmium. $^{111}$In, however, is radioactive with a half-life of 2.8 days, 
which makes using this atom to measure the mass of the electron neutrino quite a challenge.
The existence of a peak only 3 eV away from the endpoint provides new opportunities. The registration of this peak by simultaneous detection of a $\gamma$ quantum from the decay of the excited state of the cadmium nucleus and the x-ray quantum from filling the $2p_{1/2}$ vacancy in the cadmium  electron shell would allow to assert that the electron neutrino mass is less than about 3 eV provided the Q value of the decay is known with an accuracy better than 3 eV. The L2 vacancy can be filled by a radiative transition from higher levels with known energies. On the other hand, the absence of such events with sufficient statistics would imply that the electron neutrino mass is above 3 eV.
}

A noticeable contribution near the endpoint is also expected from resonances located in the non-physical region, in our case from the level of $2p_{1/2}$, if we take the reported average value of $Q_{\mathrm{EC}} = 3690$ eV. The corresponding indium energy spectrum as a function of $E - Q_{\mathrm{EC}}$ differs from that shown in Fig.~\ref{ECplot40} only near the endpoint. In its vicinity, the spectrum curve is located between the curves of the holmium and dysprosium spectra, i.e. quite high, as a result of the contribution of the subthreshold level $2p_{1/2}$.

\section{Conclusions and Outlook}
The ground-state-to-ground-state EC Q value of $^{111}$In was measured directly for the first time utilizing the double Penning trap mass spectrometer JYFLTRAP. The new Q value is 20-fold more precise and -2.57~keV lower than the adopted value from AME2020. Combined with the excitation energy from $\gamma$-ray spectroscopy, we confirm the EC transitions from $^{111}$In (ground-state) to two excited states of $^{111}$Cd with excitation energies of 866.60(6) keV and  864.8(3) keV to be energetically forbidden at a level of more than 20$\sigma$. Two other EC transitions to excitation energies of 853.94(7) keV and  855.6(10) keV are verified to be energetically positive. The EC spectrum shape of the allowed transition  $^{111}$In  (9/2$^{+}$) $\rightarrow$ $^{111}$Cd (7/2$^{+}$, 853.94(7) keV),  with a refined $Q_{\mathrm{EC}}$ value of 3.69(19) keV, was investigated with atomic Dirac-Hartree-Fock many-body calculations. We found an enhancement of event rate near the end-point area compared to $^{163}$Ho, which is so far the only nucleus used for direct neutrino mass determination. These findings unveil that it is a suitable transition for direct neutrino mass determination. The new $Q$ value of the second-forbidden unique EC transition  $^{111}$In  (9/2$^{+}$) $\rightarrow$ $^{111}$Cd (3/2$^{+}$, 855.6(10) keV)  was determined to be 2.0(10) keV. 
To further determine whether this transition  is an ultra-low $Q$ value branch, more precise excitation energy measurement is highly desired.

{
The requirement of measuring the EC Q value of $^{111}$In with highest possible accuracy is also important for determining the position of the electron level $2p_{1/2}$ relative to the endpoint of the energy spectrum. 
In the event that the level is in the physical region, new possibilities arise for measurement of the electron neutrino mass.
}


\bibliography{my-final-bib-from-jabref_titles} 

\textbf{Declaration of competing interest}

The authors declare that there are no known competing financial interests or personal relationships that could have appeared to influence the work reported in this paper.

\textbf{Acknowledgements}


We acknowledge the staff of the accelerator laboratory of University of Jyv\"askyl\"a (JYFL-ACCLAB) for providing stable online beam and J.~Jaatinen and R.~Sepp\"al\"a for preparing the production target. We thank the support by the Academy of Finland under the Finnish Centre of Excellence Programme 2012-2017 (Nuclear and Accelerator Based Physics Research at JYFL) and projects No. 306980, 312544, 275389, 284516, 295207, 314733, 318043, 327629, 320062 and 318043. The support by the EU Horizon 2020 research and innovation program under grant No. 771036 (ERC CoG MAIDEN) is acknowledged.


\bibliographystyle{model6-num-names}





\end{document}